\documentclass[a4paper]{jpconf}

\usepackage{amsmath} \usepackage{graphicx} \usepackage{amsfonts}
\usepackage{array} \usepackage{amsthm} \usepackage{bm} \usepackage{mathptmx}

\usepackage{latexsym}
\newcommand{\be}{\begin{equation}}
\newcommand{\ee}{\end{equation}}
\newcommand{\ba}{\begin{eqnarray}}
\newcommand{\ea}{\end{eqnarray}}
\newcommand{\bal}{\begin{align}}
\newcommand{\eal}{\end{align}}
\newcommand{\lb}{\label}

\newcommand{\dd}{{\rm d}}

\newcommand{\om}{\omega}
\newcommand{\al}{\alpha}

\newcommand{\ka}{\kappa}
\newcommand{\pa}{\partial}

\newcommand{\si}{\sigma}

\newcommand{\fr}{\frac}

\newcommand{\bw}{\begin{widetext}}
\newcommand{\ew}{\end{widetext}}

\newcommand{\cM}{{\cal M}}
\newcommand{\cN}{{\cal N}}

\newcommand{\cE}{{\cal E}}
\newcommand{\cF}{{\cal F}}

\begin{document}
\title{Classification of
BPS instantons in N=4 D=4 supergravity}

\author{Dmitri Gal'tsov}

\address{
 Department of Theoretical Physics, Moscow State University, 119899,
Moscow, Russia }

\ead{galtsov@phys.msu.ru}

\begin{abstract}
This talk is  based on the recent work in collaboration with M.
Azreg-A\"{\i}nou and G.~Cl\'ement \cite{acg} devoted to extremal
instantons in the one-vector truncation of the Euclidean $\cN=4,\,
D=4$ theory. Extremal solutions  satisfying the no-force condition
can be associated with null geodesic curves in the homogeneous
target space of the three-dimensional sigma model arising in
toroidal reduction of the four-dimensional theory. Here we
(preliminarily) discuss the case of two vector fields sufficient to
find all relevant metrics in the full $\cN=4,\, D=4$ theory.
Classification of instanton solutions  is given along the following
lines. The first is their possible asymptotic structure:
asymptotically locally flat (ALF), asymptotically locally Euclidean
(ALE) and ALF or ALE with the dilaton growing at infinity. The
second is the algebraic characterization of matrix generators
according to their rank and the nature of the charge vectors in an
associated Lorentzian space. Finally, solutions are distinguished by
the number of independent harmonic functions with unequal charges
(up to four).
\end{abstract}

\section{Introduction}

Instantons in vacuum Einstein gravity were subject of intense
investigations since the late seventies, which culminated in their
complete topological classification \cite{inst}. Instantons in
extended supergravities are non-vacuum and typically involve
multiplets of scalar and vector fields in four dimensions and form
fields in higher dimensions.  Non-vacuum axionic gravitational
instantons attracted attention in the late eighties in connection
with  the idea of multiverse mechanism of fixing the physical
constants \cite{axion}. These are particular solutions of the
present theory whose bosonic sector is frequently termed as the
Einstein-Maxwell-dilaton-axion (EMDA) model. All extremal instanton
solutions in the one-vector EMDA theory were found recently in
\cite{acg}. Here we make some steps toward classification of
extremal instantons in the full $\cN=4,\, D=4$ theory.

Supersymmetric solutions to the Lorentzian $\cN=4$ supergravity were
classified solving the  Killing spinor equations in
\cite{Tod:1995jf,Bellorin:2005zc}. An alternative technique to
classify BPS solutions relates to classification of null geodesics
of the target space of the sigma model obtained via dimensional
reduction of the theory along the Killing coordinate. The method was
suggested in the context of the five-dimensional Kaluza-Klein (KK)
theory by G. Clement  \cite{gc}, building on the interpretation by
Neugebauer and Kramer \cite{nk} of solutions depending on a single
potential as geodesics of the three-dimensional target space. It was
further applied in \cite{bps} to classify Lorentzian extremal
solutions of the EMDA theory and Einstein-Maxwell (EM) theory. In
two of these three cases (KK and EMDA) it was found that the matrix
generators $B$ of null geodesics split into  a nilpotent class
($B^n=0$ starting with certain integer $n$), in which cases the
matrix is degenerate ($\det B=0$), and a non-degenerate class ($\det
B\neq 0$). The solutions belonging to the first class are regular,
while the second class solutions, though still satisfy  the no-force
constraint  on asymptotic charges, generically contain
singularities. More recently similar approach  partially overlapping
with the present one was suggested as  the method of nilpotent
orbits  \cite{Boss}. The latter starts with some matrix condition
following from supersymmetry, which is generically stronger than our
condition selecting the null geodesic subspace of the target space.
In the minimal $N=2$ theory all null geodesics are nilpotent orbits,
corresponding to the Israel-Wilson-Perj\`es solutions \cite{bps},
while in the $N=4$ case there are null geodesics whose generators
are not nilpotent. These correspond  to solutions satisfying the
no-force condition, but presumably not supersymmetric.

\section{The setup}
Bosonic sector of $\cN=4, \;D=4$ supergravity  contains
 \be
g_{\mu\nu}, \quad \phi  \;\;[{\rm  dilaton }],\quad   \ka \;\;[{\rm
axion}], \quad{\rm six \;\; vector\;\; fields}\;  A_\mu.
 \ee  The theory has global  symmetries: S-duality  $SL(2,R)$
  and   $O(6)$, rotating the vector fields. The scalar fields
parametrize the coset   $SL(2,R)/SO(2)$.
\subsection{Euclidean action} Correct choice of the Euclidean action for the
axion field  follows from the positivity requirement and amounts to
starting with the three-form field:
\begin{equation}\label{acH}
    S_0=\frac1{16\pi}\int\limits_{\cM}   \left(
    -R\star 1+2  \dd\phi\wedge \star
\dd\phi+  2\e^{-4\phi}  H\wedge  \star H + 2\e^{-2\phi}
 F_n\wedge \star  F_n \right)\; - \frac1{8\pi}\int\limits_{\pa\cM}\e^{\psi/2}K\star\dd \Phi\,,
\end{equation}
 where $F_n=\dd A_n$ are the Maxwell two-forms and the sum over $n$ from one to six
is understood.  $H$ is the three-form field strength related to the
two-form potential $B$ via the relation involving the Chern-Simons
term:
$
H=\dd B-A_n\wedge F_n.
$
 The boundary $\pa\cM$ of $\cM$  is described by
$\Phi(x^{\mu})\equiv 0$, while $\e^{\psi/2}$ is a scale factor
ensuring that $\e^{\psi/2}\dd \Phi$ measures the proper distance in
a direction normal to $\pa\cM$, and $K$ is the trace of the
extrinsic curvature of $\pa\cM$ in $\cM$.

 To pass to the
pseudoscalar axion  one has to ensure the Bianchi identity for $H$:
$
  \dd \dd B=\dd (H + A\wedge F)=0\,
$ which is effected adding to the action (\ref{acH}) a new term with
the Lagrange multiplier $\ka$
 \be
S_{\kappa}= \frac1{8\pi}\int_{\cM'} \ka\; \dd (H + A_n\wedge F_n)=
\frac1{8\pi}\int_{\cM'}\ka \; (\dd H + F_n\wedge F_n)\,,
 \ee
where $\cM'$ is $\cM$ with the monopole sources of $H$  cut out.
Integrating out the three-form $H$ we obtain the bulk action  in
terms of the pseudoscalar axion
\begin{equation}\label{ac2}
    S_E= \frac1{16\pi}\int\limits_{\cM}   \left(
    -R\star 1+2  \dd\phi\wedge
\star\dd\phi- \fr12\, \e^{4\phi}  \dd \ka\wedge \star\dd\ka  +
2\e^{-2\phi}
 F\wedge  \star F + 2\ka F\wedge F\right)\,
\end{equation}
plus the boundary term. Combining the latter with the gravitational
boundary term we  get
 \be\lb{GHax}
\textsuperscript{4}S_b= \frac1{16\pi}\int\limits_{\pa\cM'}[\ka
\e^{4\phi}\star \dd\ka] -
\frac1{8\pi}\int\limits_{\pa\cM}\e^{\psi/2}[K]\star\dd \Phi,
 \ee
where the pull-back of the three-form $\star \dd\ka$ onto the
boundary $\pa\cM$ is understood. Square brackets denote the
background subtraction which is necessary to make the action finite.
Note that the bulk matter action in the form (\ref{ac2}) is not
positive definite in contrast to (\ref{acH}): the difference is
hidden in the boundary term.
\subsection{3D sigma-model}
To develop generating technique for instantons we apply dimensional
reduction to three dimensions, where the equations of motion are
equivalent to those of the sigma model on the homogeneous space of
the three-dimensional U-duality group.  The derivation of the sigma
model in the case $p=1$ was first given in \cite{emda} and
generalized to arbitrary $p$ in \cite{Gal'tsov:1996cm}. This leads
to the homogeneous space of the group $SO(2,p+2)$.  In the
particular case $p=2$ the coset has a simpler representation
$G/H=SU(2,2)/\left(SO(1,3)\times SO(1,1)\right)$
\cite{Gal'tsov:1997kp} due to isomorphism $SO(2,4)\sim SU(2,2)$.

We parametrize the four-dimensional metric as
\begin{equation}\label{an}
\dd s^2=g_{\mu\nu}\dd x^\mu \dd x^\nu=f(\dd t-\omega_i\dd
x^i)^2+\frac{1}{f}\,h_{ij}\dd x^i\dd x^j\,,
\end{equation}
where where $t$  is an Euclidean coordinate with period $\beta$, and
$f,\,\omega_i,\,h_{ij}$ are functions of $x^i$ ($i=1,2,3$).
Occasionally we will also use an exponential parametrization  of the
scale factor $f=\e^{-\psi}$.

To be able to compute the on-shell instanton action one has to keep
all boundary terms \cite{acg} which were neglected in \cite{emda,
Gal'tsov:1996cm, Gal'tsov:1997kp}  .

The Maxwell fields are parameterized by the electric $v_n$ and
magnetic $u_n$ potentials partly solving the equations of motion
\begin{align} &F_{i4}=\frac{1}{\sqrt{2}}\,\partial_iv\,,\\&
\lb{mag} \e^{-2\phi}F^{ij}-\kappa {\tilde
F}^{ij}=\frac{f}{\sqrt{2h}}\,\epsilon^{ijk}
\partial_ku\,,
\end{align}
where the index $n$ labeling different vector fields is omitted. The
rotation one-form $\om$ in the metric is dualized to the NUT
potential $\chi$:\be\lb{twist}
\partial_i\chi +v\partial_iu-u\partial_iv=-f^2h_{ij}\,
\frac{\epsilon^{jkl}}{\sqrt{h}}\,\partial_k\omega_l \ee
 (we define $\epsilon_{1234}=+1$). The resulting full bulk action is that
of the gravity-coupled three-dimensional sigma model
 \be\lb{acsig}
S_\sigma = -\frac{\beta}{16\pi}\int \dd ^3x \sqrt{h}\left({\cal R}-
G_{AB}\partial_iX^A\partial_j X^B h^{ij}\right)\,\,,
\end{equation}
where the target space variables are $\textbf{X} =
(f,\phi,v_n,\chi,\kappa,u_n)\,,$ integration is over the three-space
$\cE$ and the target space metric $\dd l^2 = G_{AB}\dd X^A \dd X^B$
reads
 \be\lb{tar4} \dd l^2 = \frac12\,f^{-2}\dd f^2 -
\frac12\,f^{-2}(\dd \chi + v_n\dd u_n - u_n\dd v_n)^2 +
f^{-1}\e^{-2\phi}\dd v_n^2- f^{-1}\e^{2\phi}(\dd u_n - \kappa \dd
v_n)^2 + 2\dd \phi^2 - \frac12\,\e^{4\phi}\dd \kappa^2\,.
 \ee
This space has the isometry group $G=SO(2,p+2)$, the same as its
Lorentzian counterpart \cite{Gal'tsov:1996cm}.  The metric
(\ref{tar4}) is the metric on the coset $G/H$, whose nature can be
uncovered from a signature argument. The Killing metric of
$so(2,p+2)$ algebra has the signature $(+2(p+1),-(p^2+3p+4)/2)$,
with plus standing for non-compact and minus for compact generators.
Since the signature of the target space is $(+(p+2),-(p+2))$, it is
clear that the isotropy subalgebra must contain $p+2$ non-compact
and $p(p+1)/2$  compact generators. Such a subalgebra of $so(2,p+2)$
is    ${\rm lie\,}(H) \sim so(1,p+1)\times so(1,1)$. We therefore
deal with the coset $SO(2,p+2)/(SO(1,p+1)\times SO(1,1))$. In the
$p=2$ case this is isomorphic to $SU(2,2)/(SO(1,3)\times SO(1,1))$.
As was argued in \cite{bps}, the maximal number of independent
harmonic functions with unequal charges is equal to the number of
independent isotropic directions in the target space. Here the
target space has $p+2$ positive and  $p+2$ negative direction, thus
the number of independent null vectors is $p+2$.

In addition to the bulk action we have a number of surface terms
resulting from three-dimensional dualizations as well as from
dimensional reduction of the four-dimensional Gibbons-Hawking-axion
term. Collecting these together, and taking care of the rescaling of
the electric and magnetic potentials, we get:
  \be\lb{btot}
S_{\rm inst} = \textsuperscript{3}S_b =
\frac{\beta}{16\pi}\int\limits_{\pa\cE} (-2[k]\ast\dd \Psi +
\ast\dd\psi_0) + \frac{\beta}{16\pi}\int\limits_{\pa\cE'}\left([\ka
\e^{4\phi}\ast \dd\ka] + 2\sqrt2u_nF_n + (\chi+u_nv_n)\cF\right) \,.
 \ee
Note that the {\em on-shell} value of the action which we are
interested in for instantons is entirely given by the boundary term
$\textsuperscript{3}S_b$ since the bulk sigma-model action vanishes
by virtue of the contracted three-dimensional Einstein equations.

Variation of the bulk action (\ref{acsig}) over $X^A$ gives the
equations of motion
 \be\lb{eqsigma}
\pa_i\left(\sqrt{h}h^{ij}G_{AB}\pa_j X^B\right)=
 \fr12\, G_{BC,A}\pa_i X^B\pa_j X^C h^{ij}\sqrt{h},
 \ee
which can be rewritten in a form explicitly covariant both with
respect to the three-space metric $h_{ij}$, and to the target space
metric $G_{AB}$
 \be\lb{consJ}
\nabla_i J^i_{A}=0\,,
 \ee
where $\nabla_i$ is the total covariant derivative involving
Christoffel symbols  both of  $h_{ij}$ and $G_{AB}$. The  currents
associated with the potentials read
 \be
J^i_{A}=h^{ij}\pa_j X^B G_{AB}\,.
 \ee
\subsection{Geodesic solutions}
Neugebauer and Kramer \cite{nk}, considering  Kaluza-Klein theory,
noticed that, if the target space coordinates  $X^A$  depend on
   $x^i$  through   the only  scalar function,
$X^A=X^A[\tau(x^i)]$,  the geodesic  curves  of the target space \be
\frac{d^2X^A}{d\tau^2}+\Gamma^A_{BC}\frac{dX^B}{d\tau}
\frac{dX^C}{d\tau}=0, \ee where $\Gamma^A_{BC}$ are Christoffel
symbols of the metric $G_{AB}$, solve the sigma-model equations of
motion, provided
 $\tau(x^i)$  is a harmonic function in three-space with the metric
$h_{ij}$: \be
\Delta\tau\equiv\frac{1}{\sqrt{h}}\partial_i\left(\sqrt{h}h^{ij}
\partial_j \tau\right)=0.
\ee Therefore certain classes of solutions can be associated with
geodesics surfaces in the target space.  Note  that no assumptions
were made here about the metric of the three-space, which is
generically
 curved.
\section{Matrix representation}
In view of the  rotational symmetry in the space of vectors, to get
all different metrics it is not sufficient to consider only one, but
it is enough to consider two vector fields. Indeed, as we will see
later, solutions can be labeled by asymptotic electric $Q_n$ and
magnetic $P_n$ charges, the metric being dependent on the three
invariants $Q^2=Q_nQ_n,\;P^2=P_nP_n$ and $PQ=Q_n P_n$. In one-vector
case $QP=0$ implies that either $Q^2=0$ or $P^2=0$. To have the
third invariant $QP$ independent of the first two, it is enough,
however, to take $p=2$: then, e.g., for $Q_1\neq 0,\; Q_2=0,\;
P_1=0,\;P_2\neq 0$ one has $QP=0$ but $Q^2\neq 0, \;P^2\neq 0$.
Using rotation in the space of vector fields, one can can always
choose this configuration as a representative of a general one. So
in what follows we will consider the case $p=2$.

To proceed, we have to introduce the matrix representation of the
coset $SU(2,2)/(SO(1,3)\times SO(1,1))$. In the Lorentz case the
corresponding coset is $SU(2,2)/(SO(2,2)\times SO(2))$, its
representation in terms of the complex matrices $4\times 4$ was
given in \cite{Gal'tsov:1997kp}. The analogous representation of the
Euclidean coset $G/H=SU(2,2)/\left(SO(1,3)\times SO(1,1)\right)$ is
given by the hermitian block matrix
 \be\lb{ME} M = \left(\begin{array}{cc}
P ^{-1}&P ^{-1}Q\\
QP ^{-1}&-P +QP ^{-1}Q
\end{array}\right)\,,
 \ee
 with $2\times 2$ hermitian blocks
 \be \lb{PQ} P = \e^{-2\phi}\left(\begin{array}{cc} f\e^{2\phi}+v_n^2& v_1-iv_2\\
 v_1+iv_2&1
\end{array}\right)\,, \quad
Q=\left(\begin{array}{cc}
v_nw_n -\chi & w_1-iw_2\\
w_1+iw_2 &-\kappa
\end{array}\right)\,,
\ee where $w_n=u_n-\ka v_n$.  In terms of these matrices the target
space metric reads \be \lb{dm}\dd l^2 = -\frac14\,\tr\left( \dd M
\dd M ^{-1}\right) = \frac12\,[(P ^{-1}\dd P )^2 - (P ^{-1}\dd
Q)^2]\,. \ee To read off the target space potentials from the matrix
$M$ it is enough to use its following two blocks:
 \ba\lb{block}
 &P^{-1} = f^{-1}\left(\begin{array}{cc} 1 & -(v_1-iv_2)\\
- (v_1+iv_2)&f\e^{2\phi}+ v_n^2
\end{array}\right)\,,& \\ & P^{-1}Q= f^{-1}\left(\begin{array}{cc}
- {\tilde \chi} &  u_1-iu_2\\ {\tilde \chi}(v_1+iv_2)
+f\e^{2\phi}(w_1+iw_2) & -\kappa f\e^{2\phi}- v_nu_n +iW
\end{array}\right)\,,&\nonumber
 \ea
 where ${\tilde \chi}=\chi+iW,\; W=v_1u_2-v_2u_1$.
\subsection{Asymptotic conditions}
We will be interested by finite action solution with vanishing
target space potentials $v_n=u_n=\ka=0$ in the asymptotic region
(specified as $r\to\infty$), while the NUT potential and the dilaton
may be growing there. If the dilaton tends to a constant value at
infinity, the asymptotic metrics can be classified as in pure
gravity \cite{inst}. These are known to be of two types. The first
includes asymptotically locally flat (ALF) solutions, with
$f(\infty) = 1,\;\chi(\infty) =0$ and the asymptotic form of the
metric
 \be\lb{metalf} \dd s^2 =(\dd
t - 2N\cos\theta\,\dd\varphi)^2 + \dd r^2 + r^2 (\dd\theta^2 +
\sin^2\theta\,\dd\varphi^2)\,,
 \ee
where $N$ is the NUT parameter. Our coset matrix $M_\infty$,
corresponding to this solution reads \be\lb{etaE}
M_{ALF}=\si_{z0}\equiv\sigma_z\otimes\sigma_0=\left(\begin{array}{cc} \sigma_0 & 0 \\
0  & -\sigma_0
\end{array}\right)\,,  \ee
where (and in what follows) we use the notation
$\si_{\mu\nu}=\si_\mu\otimes\si_\nu,\; \mu,\,\nu=0,1,2,3$ for the
direct product of the matrices $\sigma_\mu = (\sigma_0, \sigma_i)$,
with $\sigma_i$ being the Pauli matrices and $\sigma_0=1$.

The second class, with  growing $f=\chi\sim r$ at infinity,
corresponds to asymptotically locally Euclidean (ALE) solutions with
the asymptotic metric
 \be\lb{4eucl}
ds^2 = \dd \rho^2 + \rho^2 \dd \Omega_3^2,\ee where the three-sphere
 is parametrized as \be \dd \Omega_3^2=\frac14[\dd \theta^2 +
\sin^2\theta \dd \varphi^2 + (\dd \eta + \cos\theta \dd
\varphi)^2]\,,
 \ee
with the angular coordinate $\eta = t$, and the radial coordinate
$\rho = (4r)^{1/2}$. In this case $M_\infty$ is
 \be
M_{ALE}=\frac12\left(\si_{30}-\si_{33}- \si_{10}-\si_{13}\right).
 \ee
Both these asymptotic solutions satisfy source-free Euclidean
Einstein equation.

Two additional asymptotic types  correspond to ALF and ALE solutions
with dilaton growing at infinity. The ALF solutions  are related to
Lorentzian solutions with the linear dilaton asymptotic
\cite{Clement:2002mb}, while the ALE ones are dilaton-axion dressed
Eguchi-Hanson and lens spaces \cite{acg}.

\subsection{Null geodesic solutions}
In the matrix terms the sigma-model field equations (\ref{consJ})
read
 \be\lb{sigeq}
\pmb\nabla\left(M^{-1}\pmb\nabla M\right)=0\,,
 \ee
where $\pmb\nabla$ stands for the three--dimensional covariant
derivative, and the scalar product with respect to the metric
$h_{ij}$ is understood. The geodesic solutions then obey the
equation
\begin{equation} \frac{\dd}{\dd \tau}\left(M^{-1}\,\frac{\dd M}{\dd
\tau}\right)=0\,,
\end{equation}
which is first integrated by
 \be\lb{B}
M^{-1}\frac{\dd M}{\dd\tau} = B \,,
 \ee
where $B$ is a constant matrix generator of the coset. A second
integration leads to the solution to the geodesic equation in the
exponential form
\begin{equation} \label{AB}
M = M_\infty\,{\rm e}^{B\tau}\,,
\end{equation}
where we assume that $\tau(\infty)=0$.

The three--dimensional Einstein equations now read
\begin{equation} \label{ei}
{\cal R}_{ij}=-\frac{1}{4}\,\tr\left(\nabla_i M \nabla_j
M^{-1}\right)\,.
\end{equation}
The parametrisation (\ref{AB})  (\ref{ei}) to
\begin{equation}
{\cal R}_{ij}=\frac{1}{4}\,(\tr B^2)\nabla_i \tau \nabla_j \tau\,.
\end{equation}
\noindent From this expression it is clear that in the particular
case
\begin{equation} \label{null}
\tr B^2 =0
\end{equation}
the three--space is Ricci--flat. In three dimensions the Riemann
tensor is then also zero, and consequently the three--space $\cE$ is
flat. We shall assume in the following that $\cE =\mathbb{R}^3$.
From Eq. (\ref{dm}) one can see \cite{gc} that the condition
(\ref{null}) corresponds to null geodesics  of the target space
\begin{equation}\lb{ngeo}
\dd l^2=\frac{1}{4}\,(\tr B^2)\,\dd \tau^2=0\,.
\end{equation}
In terms of the above notation for $4\times 4$ matrices the eight
generators of the coset $G/H=SU(2,2)/(SO(1,3)\times SO(1,1))$ can be
chosen as \be\lb{generators} B=\left\{ {\rm lie\,}(G)\ominus{\rm
lie\,}(H)\right\}=\left\{\si_{3\mu},\; i\si_{2\mu} \right\}.\ee
\subsection{Charge vectors}
In the ALF case one assumes the following   behavior of the target
space variables at spatial infinity:
 \begin{eqnarray} \label{as}
f  \sim  1-\frac{2M}{r}\,,  \quad &&\chi  \sim    -\frac{2N}{r}\,,\nonumber\\
\phi   \sim  \frac{D}{r}\,,\quad &&  \kappa \sim \frac{2A}{r}\,, \nonumber\\
v_n \sim   \frac{\sqrt{2}Q_n}{r}\,, \quad &&
u_n\sim\frac{\sqrt{2}P_n}{r}\,.
\end{eqnarray}
Then comparing (\ref{ME}, \ref{PQ}) with (\ref{etaE}, \ref{AB}) and
using the basis (\ref{generators}), the matrix generator $B$ can be
parametrized by two vectors $\mu^\alpha,\;\nu^\alpha$ in the
four-dimensional flat space with the $SO(1,3)$ metric
$\eta_{\mu\nu}$ of the signature $(-+++)$  as follows: \be\lb{B}
B(\mu,\nu)=\nu^0\si_{30}+i\nu^i\si_{2i}+i\mu^0\si_{20}+\mu^i\si_{3i},
\ee where explicitly
 \be\lb{munu}
\mu^\al=\left(N-A,\; -\sqrt{2} Q_1,\; -\sqrt{2}
Q_2,\;M-D\right),\quad \nu^\al=\left(M+D,\; \sqrt{2} P_1,\; \sqrt{2}
P_2,\;N+A\right).
 \ee
In the space of charges the $SO(1,3)\times SO(1,1)$ global symmetry
is acting. Fixing the corresponding ``Lorentz frame'' one can
simplify matrices describing physically distinct classes of the
solutions.
\section{Classification of null geodesics}
Squaring the matrix (\ref{B}) we obtain
 \be\lb{B2}
B^2=(\mu^2-\nu^2)\si_{00}+2\si_{0i}\left(\nu^0\mu^i-\mu^0\nu^i\right)
+2i\epsilon_{ijk}\mu^i\nu^i\si_{1k},
 \ee
 where $\mu^2=\mu^\al\mu^\beta\eta_{\al\beta}$ etc. Its diagonal
 part is proportional to the difference of squared charge vectors,
 while the non-diagonal part is defined through their skew product.
\subsection{No-force condition}
To ensure the three-space to be flat, which presumably corresponds
to BPS solutions, we must impose the vanishing  trace condition on
$B^2$, which in view of (\ref{B2}) reduces to the equality of the
norms of two charge vectors $\mu^2=\nu^2$, indeed
 \be
\tr B^2=4(\mu^2-\nu^2)=0.
 \ee
 Substituting (\ref{munu}) we get the
relation between the asymptotic charges
 \begin{equation}\label{BPS}
    M^2+D^2+Q^2=N^2+A^2+P^2\,,
\end{equation}
where $Q^2=Q_nQ_n,\; P^2=P_n P_n$. This is the no-force condition in
the Euclidean case, where the mass, the dilaton charge and the
electric charges are attractive, while the NUT charge, the axion
charge and the magnetic charge are repulsive.
\subsection{Characteristic equation}
Imposing the condition (\ref{BPS}), we get for the third power of
$B$
 \be
   \frac12 B^3=\left( \mu^2\nu^0-(\mu\nu)\mu^0\right)\si_{30}+
   \left((\mu\nu)\nu^i -\nu^2\mu^i\right)\si_{3i}-i
   \left( \nu^2\mu^0-(\mu\nu)\nu^0\right)\si_{20}-i
   \left((\mu\nu)\mu^i -\mu^2\nu^i\right)\si_{2i},
   \ee
where $(\mu\nu)=\mu^\al\nu_\al$. The forth power, again with
(\ref{BPS}), is
 \be \lb{B4}
 \frac14B^4=\left( (\mu\nu)^2-\mu^2\nu^2\right)\si_{00}.
 \ee
It is easy to check that
 \be \tr B=0,\quad \tr B^3=0,
  \ee
so, together with (\ref{BPS}), one finds the following
characteristic equation for the matrix $B$
 \be\label{s2} B^4+(\det B)I=0\,,
 \ee
consistently with $B^4$ being proportional to the unit $4\times 4$
matrix.  In view of (\ref{s2}), if the matrix $B$ is degenerate,
$\det B =0$, one has $B^4=0$, so the expansion of the exponential in
(\ref{AB}) contains  only the terms up to cubic. In the
non-degenerate case the series is infinite. It turns out that in the
latter case most of the solutions contain singularities and are not
supersymmetric, so we do not discus them here.

The degeneracy condition in terms of the charge vectors (restricted
by the no-force condition) according to (\ref{B4}) is  one of the
two
 \be\lb{dege}
2 (M\pm N)(D\mp A)=(Q_n\mp P_n)^2,
 \ee
where the sum over $n$ is understood.
\subsection{Strongly degenerate case}
The  rank of the degenerate $B$ can be either two  or three. In the
first case $B^2=0$ and the coset matrix $M$ is linear in terms of
$B$:
 \be \lb{ML}
M=\eta(I+B\tau)\,.
 \ee

According to (\ref{B2}), vanishing of $B^2$, apart from (\ref{BPS})
imposes the following conditions on the charge vectors:
 \be
 \nu^0\mu^i-\mu^0\nu^i=0,\quad \mu^i\nu^j-\mu^j\nu^i=0,
 \ee
which are equivalent to vanishing of the bivector $\mu^\al\wedge
\nu^\beta$.

 This leads to different subclasses of solutions according to
whether the charge vectors $\mu^\al,\;\nu^\al$ are time-like,
space-like or null. Further details of classification of rank two
solutions in the case $p=1$ can be found in \cite{acg}. They include
solutions of all mentioned above asymptotic types.
\subsection{Weakly degenerate case}
In the case of rank three  all terms in the series expansion of $M$
up to the third are non-zero:
\begin{equation}\label{w1}
    M=\eta(I+B\tau+B^2\tau^2/2+B^3\tau^3/6)\,,
\end{equation}
 while $B^4$ and higher terms vanish by virtue of the degeneracy
 condition $\det B=0$. Now we have only two conditions on eight (for
 $p=2$)
charges: (\ref{BPS}) and one of the two in (\ref{dege}). Again this
case includes a variety of new ALF, ALE and dilatonic instantons.
\subsection{Multiple harmonic functions}
The construction (\ref{AB}) may be generalized \cite{gc,bps} to the
case of several truly independent harmonic functions
$\tau_a,\;\Delta \tau_a =0$, by replacing the exponent in (\ref{AB})
by a linear superposition
\begin{equation} \label{mupt}
 M = A \exp \left(\sum_a B_a \tau_a\right).
\end{equation}
This solves the field equations (\ref{sigeq}) provided that the
commutators $[B_a, B_b]$ commute with the $B_c$ (for the proof see
\cite{bps}):
\begin{equation} \label{dcom}
[\,[B_a, B_b], B_c] = 0 \,.
\end{equation}
The three-dimensional Einstein equations (\ref{ei}) generalize to
\begin{equation}
R_{ij} = \frac{1}{4} \,\sum_a \sum_b \tr (B_a B_b) \,\nabla_i\tau_a
\nabla_j\tau_b \,,
\end{equation}
so that the three-space is Ricci flat if the matrices $B_a$ satisfy
\begin{equation} \label{bal}
\tr (B_a B_b) = 0 \,.
\end{equation}
The number of independent harmonic functions on which an extremal
solution of the form (\ref{mupt}) may depend is limited by the
number of independent mutually orthogonal null vectors of the target
space. In the present case of Euclidean EMDA with two vector fields
this number is four. This gives a number of solutions, whose
explicit form (in the case $p=1$) can be found in \cite{acg}.

 \section{Concluding remarks}
We described general structure of the space of extremal instantons
in N=4 D=4 supergravity as null geodesics of the coset
$G/H=SU(2,2)/\left(SO(1,3)\times SO(1,1)\right)$. A number of
particular $p=1$ new solutions was given in \cite{acg}. Apart from
some simple extremal solutions, which were previously known
explicitly in the purely scalar ALE sector \cite{axion},  new scalar
ALF and ALE were found, such as dilaton-axion dressed Taub-NUT,
Eguchi-Hanson and lens-space instantons. There are also new types of
wormholes interpolating between ALF or ALE and conical ALF spaces.
All electrically and magnetically charged solutions are entirely new
except for those which were (or could be) found by euclideanization
of known Lorentzian black hole and/or IWP-type solutions, which were
rederived in the general treatment as well. The new charged ALE
solutions include, among others, purely electric solutions, as well
as purely magnetic instantons with linear dilaton asymptotics.

\section*{Acknowledgments} The author  thanks Cestmir Burdik for the
invitation to QTS7 conference in Prague and A. Strominger for useful
remarks. He is especially grateful to  M. Azreg-A\"{\i}nou and G.
Cl\'ement for fruitful and enjoying collaboration.  The work was
supported by the RFBR project 11-02-01371-a.


\section*{References}
\begin{thebibliography}{9}
\bibitem{acg}
 Azreg-A\"{\i}nou M, Cl\'ement G and  Gal'tsov D V 2011
 ``All extremal instantons in Einstein-Maxwell-dilaton-axion theory,''
  arXiv:1107.5746 [hep-th] to appear in Phys. Rev. D.

\bibitem{inst}
 Gibbons G W and  Hawking S W 1977 {\it Phys. Rev.} D  {\bf 15},
2752; S.W. Hawking S W 1977 {\it Phys. Lett.}  A {\bf 60} 81 ;
 Gibbons G W and  Hawking S W 1979
  %``Classification Of Gravitational Instanton Symmetries,''
  {\it Commun.  Math.  Phys.}  {\bf 66} 291 .

\bibitem{axion}
%\bibitem{Giddings:1988wv}
 Giddings S B and  Strominger A 1988
%``Axion Induced Topology Change In Quantum Gravity And String Theory,''
{\it Nucl.\ Phys.} B {\bf 306} 890;
%``String Wormholes,''
1989 {\it Phys.\ Lett.} B {\bf 230} 46.

%\cite{Tod:1995jf}
\bibitem{Tod:1995jf}
   Tod K P 1995
  %``More On Supercovariantly Constant Spinors,''
  {\it Class.\ Quant.\ Grav.}  {\bf 12} 1801.
  %%CITATION = CQGRD,12,1801;%%

%\cite{Bellorin:2005zc}
\bibitem{Bellorin:2005zc}
   Bellorin J and  Ort\'{\i}n T 2005
  %``All the supersymmetric configurations of N = 4, d = 4 supergravity,''
  {\it Nucl.\ Phys.}  B {\bf 726} 171;
%\cite{Meessen:2010fh}
%\cite{Meessen:2010fh}
%\bibitem{Meessen:2010fh}
   Meessen P,  Ortin T,  Vaula S 2010
  %``All the timelike supersymmetric solutions of all ungauged d=4 supergravities,''
  {\it JHEP }{\bf 1011} 072 (2010).

\bibitem{gc}
 Cl\'ement G  1986 {\it Gen. Rel. and Grav.} {\bf 18}  861; {\it
Phys. Lett.} A {\bf  118}  11.

\bibitem{nk}
 Neugebauer G and  Kramer D 1969 {\it Ann. der Physik (Leipzig)} {\bf
24} 62.

\bibitem{bps}  Cl\'ement G and  Gal'tsov D V 1996 {\it Phys. Rev.} D
{\bf 54} 6136.

\bibitem{Boss}
%\bibitem{Bossard:2009my}
%\bibitem{Bossard:2009at}
   Bossard G,  Nicolai H, K. Stelle K S 2009
  %``Universal BPS structure of stationary supergravity solutions,''
  {\it JHEP} {\bf 0907}, 003 (2009)  [arXiv:0902.4438 [hep-th]];
   Bossard G 2010
  %``The Extremal black holes of N=4 supergravity from so(8,2+n) nilpotent orbits,''
  {\it Gen.\ Rel.\ Grav.}  {\bf 42}, 539.
\bibitem{emda}
 Gal'tsov D V and  Kechkin O V 1994 {\it Phys. Rev.}   D   {\bf 50} 7394;
1995 {\it Phys. Lett.} B {\bf361}, 52 (1995); 1996 {\it Phys. Rev.}
D {\bf 54}  1656;
%\bibitem{g}
 Gal'tsov D V 1995 {\it Phys. Rev. Lett.} {\bf 74} 2863.


\bibitem{Gal'tsov:1996cm}
   Gal'tsov D V and  Letelier P S 1997
  %``Ehlers-Harrison transformations and black holes in dilaton-axion  gravity
  %with multiple vector fields,''
  {\it Phys.\ Rev.}  D {\bf 55}  3580.
  %%CITATION = PHRVA,D55,3580;%%

%\cite{Gal'tsov:1997kp}
\bibitem{Gal'tsov:1997kp}
   Gal'tsov D V  and  Sharakin S A 1997
  %``Matrix Ernst potentials for EMDA with multiple vector fields,''
  {\it Phys.\ Lett.} B {\bf 399}  250.
  %%CITATION = PHLTA,B399,250;%%
%\cite{Clement:2002mb}
\bibitem{Clement:2002mb}
 Cl\'ement G, Gal'tsov D V and Leygnac C 2003
  %``Linear dilaton black holes,''
  {\it Phys.\ Rev.}  D {\bf 67} 024012
   .
  %%CITATION = PHRVA,D67,024012;%%

\end{thebibliography}
\end{document}